%
%
%
%
%
%
\def\ni{\noindent}
\def\ea{{\it et al\/}}
\def\ie{{\it i.e.\ }}

\def\bi{\langle n_1\, j_1 \, \epsilon_1 |}
\def\bi1{\langle 1|}
\def\bd{{|n_2\,j_2\,\epsilon_2\rangle}}
\def\bd2{{| 2\rangle}}

%
 \newbox\Ancha
 \def\gros#1{{\setbox\Ancha=\hbox{$#1$}
   \kern-.025em\copy\Ancha\kern-\wd\Ancha
   \kern.05em\copy\Ancha\kern-\wd\Ancha
   \kern-.025em\raise.0433em\box\Ancha}}
%

\input iopppt
\pptstyle
\jl{2}

\paper{ Recurrence relations for   relativistic  two centre  matrix elements.}

\author{A C Ilarraza-Lomel\'i,\dag  R P Mart\'inez-y-Romero,\ddag\footnote{\S}{E-mail rodolfo@dirac.fciencias.unam.mx  } H N N\'u\~nez-Y\'epez,$^+$ \footnote{$\|$}{E-mail nyhn@xanum.uam.mx}  A L Salas-Brito,\dag \footnote{\P}{Corresponding author, e-mail: asb@correo.azc.uam.mx } and M N Vald\'es-Mart\'inez,\dag}[A C Ilarraza-Lomel\'i \etal\ ]

\address{\dag Laboratorio de Sistemas Din\'amicos, Departamento de Ciencias B\'asicas, Universidad Aut\'onoma Metropolitana, Unidad Azcapotzalco,  Apartado Postal 21-267,  Coyoac\'an, Distrito Federal, C P 04000, M\'exico }

\address{\ddag Facultad de Ciencias, Universidad Nacional Aut\'onoma de M\'exico, Apartado Postal 50-542, Mexico City, Distrito Federal, CP 04510, Mexico }

\address{$^+$ Departamento de F\'isica, Universidad Aut\'onoma Metropolitana, Unidad Iztapalapa, Apartado Postal 55-534, Iztapalapa, Distrito Federal, C P 09340, M\'exico}

\abs 
Recursion formulae are derived for the calculation of  two  centre matrix elements of a radial function  in  relativistic quantum mechanics.  The recursions are obtained  between  not necessarily diagonal radial eigensates using arbitrary radial potentials and any radial functions.  The only restriction is that the potentials have to share a  common minimum. Among other things, the relations so obtained  can help in  evaluating  relativistic corrections to transition probabilities between atomic Rydberg states. \endabs

\submitted 
\date
\pacs{31.30.Jv, 03.65.Pm, 03.20.+i}
\section {Introduction}
The bridge between quantum calculations and experimental results is made at the level of  matrix elements of the appropriate operators (Wong and Yeh 1983a,b De Lange and Raab 1991, Dobrovolska and Tutik  1999, Quiney \etal\ 1997, Elboudali and Joulakian 2001, Ilarraza-Lomel\'i \etal\ 2002). In  atomic and  molecular physics this usually means matrix elements of powers of a radial coordinate between  states of the system at hand (Blanchard 1974, Brack 1983, Bessis \etal\ 1985, Nana Engo \etal\ 1997, Owono Owono \etal\ 2002, West 2001). Matrix elements of more general radial functions are also very useful (Avram and Dr\u{a}g\u{a}nescu 1997, Charro and Martin 2002, Nagy and V\'egh 1992). In nonrelativistic quantum mechanics, the  importance of  hypervirial results,   and other related   techniques, follows from this fact,  since  the task of calculating  matrix elements is simplified using such techniques (Morales 1987, Fern\'andez and Castro 1987, 1996, De Lange and Raab 1991, N\'u\~nez-Y\'epez \etal\ 1995).   These techniques are also important for atomic physics in the relativistic realm (Mart\'inez-y-Romero \etal\ 2001). This is  especially important at present given the precision attained in atomic physics  experiments using synchroton radiation and other state-of-the-art sources of photons (Schippers \etal\ 2002,  2003, M\"uller \etal\ 2002, Aguilar \etal\ 2003).

In this work we exhibit that hypervirial-like formulas and other operator techniques suffice to obtain  recurrence relations  between relativistic matrix elements of an arbitrary  radial function  between states  corresponding to two different arbitrary potentials. Such relation  generalizes other  recurrence relations between  relativistic hydrogenic matrix elements of powers of $r$ and of $\beta r$  [$ \beta$ is a Dirac matrix, see equation (2)] we have recently calculated (Ilarraza-Lomel\'i \etal\ 2002, Mart\'inez-y-Romero \etal\ 2000, 2001, 2002). In subsections 2.1, 2.2 and 2.3 of the paper we obtain the four recursion relations  between the matrix elements  of an arbitrary radial function, $f(r)$, with the corresponding  matrix elements of its first and  second derivatives, taken between eigenstates of different potential  functions, $V_1(r)$ and $V_2(r)$, with the corresponding  matrix elements of its first and  second derivatives. These generalized recurrence relations  follow from a hypervirial-like result together with  operator algebra computations.  The  recursions so obtained  may be  useful  for studying radiative transitions in Rydberg atoms, in analysing atomic photorecombination and photoionization processes,   for calculating  relativistic corrections to  ionic oscillator strengths,   or in analysing impact ionization or  vibrational transitions in molecules --- albeit in the last two cases in a  crude manner --- (Schippers \etal\ 2002, Owono Owono \etal\ 2002, Semenov and Cherepkov 2003, Weck \etal\ 2001, Bauche-Arnoult \etal\ 1989).   Section 3 contains the conclusions and other final remarks.

\section{Relations for two centre integrals in relativistic quantum mechanics}

General hypervirial results,  the virial theorem,  and other algebraic and operator techniques  have been always very useful for calculating matrix elements in quantum mechanics. Nonetheless they have been little used in relativistic situations, see, however, (Brack 1983, Lucha and Sch\"oberl 1990). We want to show in this paper that they have  similar importance for obtaining recursion relations between matrix elements of radial functions between eigenstates of different radial potentials in relativistic quantum mechanics.   Thus,  let us  consider two  radial Dirac Hamiltonians  with two possibly different radial  potentials (each behaving as the temporal component of a 4-vector) $V_1(r)$ and $V_2(r)$. We  further assume that these potentials have the same equilibrium position which, furthermore, is assumed coincident with the origin of coordinates. That is,  the recurrence relations correspond to the so-called unshifted case. The potentials could be regarded  as describing the electronic configurations in processes involving electronic rearrangements in atomic transitions. For example, any transition  to an  autoionizing state studied in the central field approximation where the electron motion is solution of the Dirac equation with an effective central potential created by a $k$ or a $k-x$ electron ion core; or any other atomic process involving highly excited electrons which needs to be studied using multichannel spectroscopy or quantum deffect theory (Owono Owono \etal\ 2002, Mannervik \etal\ 1998, Xiao-Min \etal\ 1994, Aymar \etal\ 1996).  

The Hamiltonians can be written as (Mart\'inez-y-Romero 2000) 

$$ \eqalign{H_1=&c\alpha_r[p_r-\i \beta\epsilon_1 (j_1+1/2)/r] +M_1 \beta c^2+V_1(r),\cr
H_2=&c\alpha_r[p_r-\i \beta\epsilon_2 (j_2+1/2)/r] +M_2\beta c^2+V_2(r),
} \eqno(1)$$

\noindent  in writing equation (1) we   assume   $M_1\neq M_2$  since it is  convenient for the calculations we  perform for deriving the recurrence relations.  The eigenstates of these radial Hamiltonians correspond to a definite value of the total  angular momentum (orbital plus spin) ${\bf J}= {\bf L} +{\bf S}$ and of the quantum number $\epsilon=(-1)^{j+l-1/2}$, where, respectively, $j$ is the total, $l$  the orbital, and $s$   the spin angular momentum quantum numbers, the operators

$$\beta=\pmatrix{1&0\cr0& -1}\quad \hbox{ and } \quad \alpha_r= \pmatrix{0& -1\cr-1& 0}, \eqno(2)$$ 

\noindent are Dirac matrices in  the two-dimensional representation appropriate for the radial problem we are discussing, and the subscript simply distinguishes between the systems pertaining to the different potentials. Each Dirac equation can  be written as

$$ H_k \psi_k(r)=E_k \psi_k(r)\eqno(3) $$

\noindent where the energy eigenvalues $E_{n_kj_ks_k} \equiv  E_k$ and the corresponding eigenfunctions $\psi_{n_kj_ks_k}(r)\equiv \psi_k(r)$ are assumed known. In this work we use atomic units ($\hbar=m_e=e=1$) so the masses in (1) are given in units of the electron mass. The radial eigenstates $\psi_k(r)$ are the purely radial spinors (Drake 1996, Greiner 1991)

$$ \psi_k(r)\equiv \psi_{n_k j_k \epsilon_k}(r) =
 {1\over r}\left( \matrix{F_{n_k j_k \epsilon_k}(r)\cr \i G_{n j \epsilon_k}(r)}\right); 
\eqno(4) $$

\ni $F$ and $G$ are, respectively, the big and the small components of the energy eigenstate. In the following we employ, as we did  in equation(3),  the shorthand $k$  for describing the whole set of quantum numbers $ n_k j_k \epsilon_k$. We also use the ket notation for the relativistic eigenstates $|n_k l_k \epsilon_k\rangle \equiv |k\rangle$. 

 \subsection{The first recurrence relation for the case of two centres}

Taking the difference between the radial Hamiltonians $H_1$ and $H_2$ in (1),  we obtain

$$ H_1= H_2 + \i c\alpha_r\beta {\Delta^-\over 2r}-c^2 \beta M^- - \left(V_2(r)-V_1(r) \right). \eqno(5)$$ 

\noindent  where $M^{\pm}\equiv M_2 \pm M_1$, and $\Delta^\pm \equiv \epsilon_2(2j_2 + 1) \pm \epsilon_1(2j_1 + 1)$. On employing  (1) again, we  can directly evaluate the commutator

$$ [H_1, f(r)]=-\i c\alpha_r {df(r)\over dr} \eqno(6)$$

\noindent where $f(r)$ is an arbitrary  radial function and $[H, f(r)]$ stands for the commutator between $H$ and $f(r)$. We can calculate this commutator again, but now using equation (5), to get the alternative form

$$ \eqalign{[H_1, f(r)]= H_2 f(r)-f(r) H_1 + \left(\i c\alpha_r\beta 
{\Delta^-\over 2r} -c^2 \beta M^- - V^-    \right)f(r).} \eqno(7)$$

It is now  simple   to  obtain, from equations (6)  and (7), the relation

$$ \hskip - 1. cm (E_{2}-E_{1})
\langle 2|f|1\rangle=  \langle 2| \left(c^2 \beta M^- + V^-\right)f |1\rangle  -
 \i c \langle 2 | \alpha_r\left( f'+\beta {\Delta^-\over 2r}f\right)|1\rangle ; 
\eqno(8) $$

\ni where we have additionally taken matrix elements between the eigenstates (4), we use the notation  $\langle 1|\equiv\langle n_1\, j_1 \, \epsilon_1 |$ and $|n_2\, j_2 \, \epsilon_2\rangle\equiv\bd2 $, and we have defined 

$$V^\pm \equiv V_2(r)\pm V_1(r). \eqno(9)$$ 

 Equation (8) may lead to recursions between relativistic matrix elements of radial functions between hydrogenic states (Mart\'inez-y-Romero \ea\ 2003, Ilarraza-Lomel\'i \ea\ 2002, Mart\'inez-y-Romero  \ea\ 2002, Mart\'inez-y-Romero \ea\ 2001) and generalizes a nonrelativistic one useful for similar purposes (N\'u\~nez-Y\'epez \ea\ 1995). Equation (8) is an exact relation for the calculation of any $f(r)$ two-centre matrix elements in relativistic quantum mechanics. Taking the potentials as equal, \ie\ $V_1(r)=V_2(r)$, we recover a relation which has proved  useful for obtaining recurrence relations between  atomic matrix elements in relativistic quantum mechanics (Ilarraza-Lomel\'i \etal\ 2002, equation (6); Mart\'inez-y-Romero \etal\ 2001, equation (24)).

 Albeit exact, equation (8) is not entirely convenient  due to the presence of the operator $\alpha_r \beta$. To get rid of this factor, we found it convenient to deal directly with  operator relations and not with the matrix elements themselves. The matrix elements will be evaluated at the end of the operator calculations.

Let us first establish that

$$ H_2 f-f H_1=\left(c^2 \beta M^-+ V^-\right)f - \i c\alpha_r \left( f' + \beta f {\Delta^- \over 2r} \right),
\eqno (10)$$

\noindent notice that equation (8) above can be  obtained from (10) just by taking matrix elements.  The following result is also easily established  

$$ H_2 f+f H_1=\left(c^2 \beta M^++ V^+\right)f - \i c\alpha_r \left( 2f{d\over dr} + f' + {2f\over r} + \beta f {\Delta^+ \over 2r} \right).
\eqno (11)$$

\noindent Then, it can be seen that

$$ \hskip -2.7 cm -\i c\left( H_2 \alpha_r f  +  \alpha_r f H_1\right)= \i c\alpha_r \left(c^2 \beta M^- - V^+\right) f -c^2\left( 2f {d\over dr}+ f' +{2f\over r} - \beta f {\Delta^-\over 2r}\right), 
\eqno(12)$$

\noindent and that 

$$ \hskip -2.2cm 
H_2 fV^- - fV^- H_1= \left(c^2 \beta M^- + V^- \right) V^-f
 -\i c\alpha_r \left(V^- f' + {dV^-\over dr} f + \beta f V^- {\Delta^-\over 2r}\right).
\eqno(13)$$

\noindent  The next relation is also readily apparent

$$ \hskip -1cm 
-\i c\left[ H_2 \alpha_r \beta {f\over r}  +  \alpha_r \beta {f\over r} H_1\right]= -\i c\alpha_r \beta V^+ {f\over r} 
-c^2\left[ \beta\left( {f'\over r}-{f\over r^2} \right)-{\Delta^+\over 2r}{f\over r}\right]. 
\eqno(14)$$

\noindent To further proceed, let us  define $\psi(r)\equiv H_2f(r)+f(r)H_1$, and  evaluate

$$ \hskip -1cm
\eqalign {
H_2&\psi-\psi H_1=c^2\beta {\Delta^+\over 2r}f'+c^2 \left({\Delta^-\over 2r}\right)^2f+\left(c^2\beta M^- + V^-\right)^2f \cr 
-&c^2f'' -c^2\beta{\Delta^-\over 2r} \left(2f{d\over dr} + f' + {f\over r}\right)\cr 
-& \i c\alpha_r\left[ \left(f'+\beta f{\Delta^-\over 2r}\right)\left(V^--c^2\beta M^+ \right) +
c^2\beta M^- \left(2f{d\over dr} +f'+{2f\over r}\right) 
\right. \cr 
+& \left. V^-f' +{dV \over dr}f+c^2M^-{\Delta^+\over 2r}f + V^- {\Delta^-\over 2r}\beta f
\right].
}\eqno (15) $$

\noindent Given these last expressions [equations (10)--(15)], it is relatively simple to obtain, from (10), 

$$ -\i c \alpha_r\left( f'+\beta f{\Delta^-\over 2r}\right)= \left(H_2 f-f H_1\right) - \left(c^2\beta M^- + V^-\right)f, 
\eqno (16) $$

\noindent and from (11),

$$ \hskip -2.5cm 
- \i c\alpha_r \beta \left( 2f{d\over dr} + f' + {2f\over r}\right) = \left( H_2 \beta f+\beta f H_1 \right) - \left(c^2\beta M^+ + V^+\right)\beta f + \i c\alpha_r {\Delta^+ \over 2r} f. 
\eqno (17) $$ 

\noindent From equation (12) we obtain

$$ \hskip -1 cm 
\eqalign {
-c^2\beta {\Delta^- \over 2r}\left( 2f {d\over dr}+ f' +{2f\over r}\right)=&  
-\i c\left( H_2 \alpha_r \beta {\Delta^- \over 2r}f  +  \alpha_r \beta {\Delta^- \over 2r}f H_1\right) \cr 
-& \i c\alpha_r \beta {\Delta^- \over 2r}\left(c^2\beta M^- - V^+\right)f -c^2 f \left({\Delta^-\over 2r}\right)^2.
}\eqno(18)$$

\noindent From equation (13) we obtain

$$ \hskip -1cm 
\eqalign {
-\i c\alpha_r \left(V^- f' + {dV^-\over dr} f \right) =& \left( H_2 fV^- - fV^- H_1\right)- 
 \left(c^2 \beta M^- + V^- \right) V^-f \cr 
+& \i c\alpha_r V^- {\Delta^-\over 2r}\beta f.
} \eqno(19)$$

\noindent  Substituting equations (16), (17), (18), and (19), into equation (15), we  get  

$$ \eqalign {
H_2&\psi -\psi H_1=-c^2\left(f''-\beta f'{\Delta^+\over 2r}\right) + c^2{\Delta^-\over 2r^2}\beta f + \left(M^-\right)^2c^4f 
 \cr 
-& c^2M^+\left(H_2\beta f-\beta f H_1 \right) + c^2M^+V^-\beta f + c^2M^-\left(H_2\beta f+\beta f H_1 \right) \cr 
-& c^2M^-V^+\beta f + V^- \left[\right.  2\left(H_2 f-f H_1\right)-V^-\left.\right] \cr 
-& \i c\left(H_2 \alpha_r \beta {\Delta^-\over 2r}f + \alpha_r \beta {\Delta^-\over 2r}f H_1 \right) - \i c\alpha_r{\Delta^-\over 2r} \left(c^2 \beta M^- - V^+\right)\beta f.
} 
\eqno (20) $$

\noindent The terms which include the operator $\i c\alpha_r$ can be also obtained  from (14), thus 
we get the expression

$$ \eqalign { 
-\i c\left(H_2 \alpha_r \beta {\Delta^-\over 2r}f + \alpha_r \beta {\Delta^-\over 2r}f H_1 \right) -& \i c\alpha_r{\Delta^-\over 2r} \left(c^2 \beta M^- - V^+\right)\beta f = \cr 
-&c^2 {\Delta^-\over 2r} \left[\beta \left(f'-{f\over r}\right)-{\Delta^+\over 2r}f\right].
} \eqno (21) $$ 

\noindent  Susbtituting now equation (21) into (20),  it easily yields  

$$ \hskip -1cm \eqalign {
H_2&\psi -\psi H_1=-c^2\left(f''-\beta f'{\Delta^+\over 2r}\right) + 2c^2{\Delta^-\over 2r^2}\beta f + \left(M^-\right)^2c^4f 
 \cr 
-& c^2M^+\left(H_2\beta f-\beta f H_1 \right) + c^2M^+V^-\beta f + c^2M^-\left(H_2\beta f+\beta f H_1 \right) \cr 
-& c^2M^-V^+\beta f + V^- \left[\right.  2\left(H_2 f-f H_1\right)-V^-\left.\right] 
- c^2 {\Delta^-\over 2r} \left(\beta f'-{\Delta^+\over 2r}f\right).
} 
\eqno (22) $$ 

Evaluating the matrix elements between the Dirac eigenstates $\langle 2|$ and $|1\rangle$ and rearranging, we finally obtain the relation

$$ \hskip -2.6cm \eqalign{ 
& a_0\langle2|f|1\rangle + a_{2}\langle2|{f\over r^2}|1\rangle - 2 E^-\langle2|{V^- f}|1\rangle + \langle2|{\left(V^-\right)^2 f}|1\rangle 
+ c^2 \langle2|f''|1\rangle \cr 
= b_0\langle2|\beta f|1\rangle &+ b_{1}\langle2|\beta {f\over r^2}|1\rangle - c^2 M^-
\langle2| V^+ \beta f|1\rangle + c^2 M^+ \langle2| V^- \beta f|1\rangle 
+ b_{4}\langle2|\beta {f' \over r}|1\rangle, 
} \eqno (23) $$ 

\noindent where 

$$ \eqalign { 
a_0 =& \left(E^-\right)^2 - \left(c^2 M^- \right)^2 \cr 
a_{2} =& -{c^2 \over 4} \Delta^- \Delta^+ \cr 
b_0 =& c^2 \left(M^- E^+ - M^+ E^- \right) \cr 
b_{1} =& c^2 \Delta^- \cr 
b_{4} =& {c^2\over 2} \left(\Delta^+-\Delta^-\right)
} \eqno (24) $$ 

\ni This is the first relation between matrix elements of an arbitrary radial function $f(r)$ between eigenstates of two different potentials as a function of the eigenenergies in  relativistic quantum mechanics.

\subsection{The second two-centre recurrence relation}

Given that the radial  eigenstates have two components in relativistic quantum mechanics,  it should be clear that we need more relations. To obtain such second equation, let us evaluate  the following operator identity --- \ie\ again we  calculate with the basic operators before any matrix element is taken.

$$ \eqalign{ H_2&  fV^-  +  fV^- H_1= \left(c^2 \beta M^+ + V^+ \right) V^-f\cr
 -&\i c\alpha_r \left(2 V^-f {d \over dr}+V^- f' + {dV^-\over dr} f +2 V^- {f\over r}+ \beta V^- f {\Delta^+\over 2r}\right).
 }\eqno(25)$$

\noindent Using again the definition $\psi(r)\equiv H_2f(r)-f(r)H_1$, we  obtain

$$ \hskip -2.5cm    
\eqalign{ H_2&\psi+\psi H_1= c^2 f {\Delta^-\Delta^+\over 4r^2}-\i c\alpha_r\left(f'+\beta f {\Delta^-\over 2r}\right)\left(V^+ - c^2 \beta M^- \right)\cr
+& \left(c^2 \beta M^+ + V^+ \right) \left(c^2 \beta M^- + V^-\right)f
- c^2 \left( 2f'{d\over dr} + f'' + 2 {f'\over r} - \beta f {\Delta^-\over 2r^2}\right)\cr
-& \i c\alpha_r \left(2 V^-f {d \over dr}+V^- f' + {dV^-\over dr} f +2 V^- {f\over r}+ \beta V^- f {\Delta^+\over 2r} + c^2 \beta M^- f' + c^2 M^- {\Delta^- \over 2r}f\right).
} \eqno(26) $$

 The calculations required for getting to the recurrence relation are similar to that used in the last section. So, from equation (10) we obtain

$$ \hskip -2.2cm 
-\i c \alpha_r c^2 \beta M^- \left( f'+\beta f{\Delta^-\over 2r}\right)= c^2 M^- \left(H_2 f-f H_1\right) - c^2 M^- \left(c^2\beta M^- + V^-\right)\beta f, 
\eqno (27) $$

\noindent from (12) we get

$$ \hskip -1 cm 
\eqalign {
-c^2 \left( 2f' {d\over dr}+ f'' +{2f'\over r}\right)=&  
-\i c\left( H_2 \alpha_r f'  +  \alpha_r f' H_1\right) \cr 
-& \i c\alpha_r \left(c^2\beta M^- - V^+\right)f' -c^2 \beta f' {\Delta^-\over 2r},
}\eqno(28)$$

\noindent and from equation  (25),  we obtain

$$ \hskip -1cm \eqalign { 
-\i c\alpha_r \left(2 V^-f {d \over dr}+V^- f'\right. +& \left. {dV^-\over dr} f +2 V^- {f\over r}\right) = \left(H_2 fV^-+fV^- H_1\right) \cr 
+& \i c\alpha_r V^- \beta f {\Delta^+\over 2r} - \left(c^2 \beta M^+ - V^+ \right) V^-f.
 }\eqno(29)$$

Using equations (16), (27), (28) and (29), in equation (26), it yields

$$ \eqalign{ 
H_2\psi+\psi H_1 =& -c^2 \beta {\Delta^-\over 2r} \left(f'-{f\over r}\right) + c^2 {\Delta^+\over 2r} {\Delta^-\over 2r}f - V^-V^+f \cr   
+& M^- M^+ c^4 f + V^+ \left(H_2 f-f H_1\right) + V^- \left(H_2 f+f H_1\right)\cr 
-& \i c \left(H_2 \alpha_r f'+\alpha_r f' H_1\right) -\i c \alpha_r \left(c^2 \beta M^- - V^+\right)f'.
} \eqno (30) $$ 

\noindent Again, the last two terms of (30) can be obtained from (14), and combining these with 

$$ \eqalign { 
-\i c\left(H_2 \alpha_r f' + \alpha_r f' H_1 \right) -& \i c \alpha_r \left(c^2 \beta M^- - V^+\right) f' = \cr 
-& c^2 \left(f''-{f'\over r}\right) + c^2{\Delta^+\over 2r}\beta f'.
} \eqno (31) $$ 

\noindent  we get  

$$ \eqalign{ 
H_2\psi+\psi H_1 =& -c^2 \beta {\Delta^-\over 2r} \left(f'-{f\over r}\right) + c^2 {\Delta^+\over 2r} {\Delta^-\over 2r}f - V^-V^+f \cr   
+& M^- M^+ c^4 f + V^+ \left(H_2 f-f H_1\right) + V^- \left(H_2 f+f H_1\right) \cr 
-& c^2 \left(f''-{f'\over r}\right) + c^2{\Delta^+\over 2r}\beta f',
} \eqno (32) $$ 

Taking matrix elements, we obtain the recurrence relation  

$$ \hskip -.6cm \eqalign {
c_0\langle2|f|1\rangle &+ a_{2}\langle2| {f\over r^2}|1\rangle - E^+ \langle2| V^- f|1\rangle - E^- \langle2| V^+ f|1\rangle + \langle2| V^+ V^- f|1\rangle \cr 
&- c^2 \langle2|{f'\over r}|1\rangle + c^2 \langle2|f''|1\rangle =  
{b_{2}\over 2}\langle2|\beta{f\over r^2}|1\rangle + b_{4}\langle2| \beta {f'\over r} |1\rangle, 
} \eqno (33) $$ 

\noindent where the only newly defined  coefficient is

$$  
c_0 = E^+ E^- - c^4 M^+ M^-.
 \eqno (34) $$

\subsection{The third and fourth two-centre recurrence relations}

To get the third recurrence relation,  we have to substitute, in a similar way to what we have done for getting the previous two relations, equations (10), (12), (14) and (25) into (26) and, after some juggling with the resulting terms, and taking  matrix elements, we finally obtain the third recurrence relation

$$ e_0\langle2|f|1\rangle = g_0\langle2|\beta f|1\rangle - \langle 2| \left( V^+-V^-\right)\beta f|1\rangle, 
\eqno (35) $$ 

\noindent  where 

$$ \eqalign {
e_0 &= c^2 \left( M^+ - M^- \right) \cr 
g_0 &= E^+ - E^-. 
} \eqno (36) $$

\ni This is a very simple equation that, besides, allows writing the matrix elements of $f$ in terms of those of $\beta f$. To take advantage of this fact, substitute equation (35) into (23) to obtain a new relation 

$$ \eqalign {
&A_0 \langle2|\beta f|1\rangle + A_1\langle2|\beta {f\over r^2}|1\rangle + A_2\langle2|V^- \beta f|1\rangle + A_3\langle2|\left(V^-\right)^2 \beta f|1\rangle \cr 
&+ \langle2|\left(V^-\right)^3 \beta f|1\rangle + A_5\langle2|V^+ \beta f|1\rangle + 2E^-\langle2|V^-V^+\beta f|1\rangle \cr 
&- \langle2|\left(V^-\right)^2 V^+ \beta f|1\rangle + a_{2}\langle2|\left(V^+-V^-\right)\beta {f \over r^2}|1\rangle = \cr 
&A_9\langle2|\beta {f' \over r}|1\rangle + c^2g_0\langle2|\beta f''|1\rangle - c^2 \langle2|\left(V^+-V^-\right)\beta f''|1\rangle. 
} \eqno (37) $$

\noindent  where the newly defined coefficients are

$$ \hskip -1.5cm \eqalign { 
A_0 &= \left(E^-\right)^2 \left(E^+-E^-\right) + c^2 E^- \left[ \left(M^-\right)^2 + \left(M^+\right)^2\right] - c^4 M^+ M^- \left(E^++E^-\right) \cr 
A_1 &= -{c^2 \over 4} \left(E^+-E^-\right) \Delta^+\Delta^- - c^4 \Delta^- \left(M^+-M^-\right) \cr 
A_2 &= -2E^- \left(E^+-E^-\right) + \left(E^-\right)^2 - \left(c^2 M^-\right)^2 - c^4 M^+ \left(M^+-M^-\right) \cr 
A_3 &= E^+ - 3 E^- \cr 
A_5 &= c^4 M^+ M^- - \left(E^-\right)^2 \cr  
A_9 &= {c^4 \over 2} \left(M^+-M^-\right) \left(\Delta^+ - \Delta^-\right).   
} \eqno (38) $$ 

\ni  Equation (37) is the fourth  recurrence  relation for the calculation of relativistic $f(r)$ two centre matrix elements in terms of the energy eigenvalues of the intervening potentials.  Notice that, at difference of the previous relations [equations (23), (33) and (35)], equation (37) just relates  among themselves matrix elements of  $\beta f$ and its derivatives times a certain function of $r$.

\section{Conclusions} 
We have obtained recurrence relations for the calculation of two-centre matrix elements of a radial function between states of two different radial potentials. The obtained recursions are given  in the most general case of an arbitrary function taken between any non necessarily diagonal radial eigenstates of two  radial potentials. These recursion relations have, as particular cases,   recursions between one-centre integrals or, in other particular cases, between overlap and one centre   integrals in Dirac relativistic quantum mechanics. We expect the obtained   recursions, together with the previous  one-centre relations we have obtained (Ilarraza-Lomel\'i \etal\ 2002, Mart\'inez-y-Romero \etal\ 2001, 2002), to be  useful in atomic or molecular physics  calculations as they may simplify  calculation  in the range of applicability of Dirac's relativistic quantum mechanics (Bang and Hansteen 2002, Moss 1972). For most uses of the relations we first have to set $M_1=M_2$, \ie\ $M^-=0$ and $M^+= 2$ ---if the particles are electrons; since the use of unequal masses is just a recourse  of our calculational method. 

From a practical angle, there is little that can be done for the analytical evaluation of two centre integrals of atomic physics interest beyond the Coulomb and the few similarly exactly solvable potentials. However, there are numerical methods that, after being adapted to relativistic conditions,  can provide the crucial ``seed'' results needed for the systematic use of the recurrence relations obtained here (Chen \etal\ 1993). Our results can be also useful in the so-called perturbation theory of relativistic corrections, in relativistic quantum deffect calculations, and for the relativistic extension of the calculations of exchange integrals using Slater orbitals or Coulomb-Dirac wave functions (Owono Owono \etal\ 2002, Bang and Hasteen 2002, Charro \etal\ 2001, Kuang and Lin 1996, Rutkowski 1996). It is also possible that our relations could be applied to a generalization of an approximate two-centre technique used for studying electron-impact ionization of simple molecules (Weck \etal\ 2001).

On the other hand, our results can be also of interest in nuclear studies. Since  the 3D Woods-Saxon potential, used in the Dirac equation for describing the interaction of a nucleon with a heavy nucleus, has been explicitly solved recently and its eigenfuctions expressed in terms of hypergeometric functions (Jian-You \etal\ 2002), so it has the features needed for the direct use of our recurrence relations.

\ack  This work has been partially supported by PAPIIT-UNAM (grant 108302). We acknowledge with thanks the comments of  C Cisneros and I \'Alvarez.  We want to thank also the friendly support of Dua and Danna Castillo-V\'azquez, and the cheerful enthusiasm of G A Inti, F A Maya, P M Schwartz, A S Ubo, G Sieriy, M Chiornaya, P A Koshka, G D Abdul, and D Gorbe.\par

\references
\refjl {Aguilar A, West J B, Phaneuf R A, Brooks R L, Folkmann F, Kjeldsen H, Bozek J D, Schlachter A S, and Cisneros C 2003} {Phys. Rev. A} {67} {012701}

\refjl {Avram N M and Dr\u{a}g\u{a}nescu Gh E  1997} {Int. J. Quantum Chem.} {65} {655}

\refjl {Aymar M, Greene C H, Luc-Koenig E 1996} {Rev. Mod. Phys.} {68} {1015}

\refjl {Bang J M and Hansteen J M 2002} {J. Phys. B: At. Mol. Opt. Phys.} {35} {3979}

\refjl {Bauche-Arnould C, Bauche J, Luc-Koenig E, Wyart J-F, More R M, Chenais-Popovics C, Gauthier J-C, Geindre J-P, and Tragin N 1989} {Phys. Rev. A} {39} {1053}

\refjl {Bessis N, Bessis G, and Roux D 1985} {Phys. Rev. A } {32} {2044}

\refjl {Blanchard P 1974} {J. Phys. B: At. Mol. Opt. Phys.} {7} {1993}

\refjl{Brack M 1983} {Phys. Rev. D } {27} {1950}  

\refjl{Charro E, L\'opez-Ferrero S, Mart\'in I 2001} {J. Phys. B: At. Mol. Opt. Phys.} {34} {4243}

\refjl {Charro E and Martin  I 2002} {J. Phys. B: At. Mol. Opt. Phys.} {35} {3227}

\refjl {Chen Z, Bessis D and Msezane A Z 1993} {Phys. Rev. A} {47} {4756}

\refbk{De Lange O L and Raab R E 1991} {Operator Methods in Quantum Mechanics} {(Oxford: Clarendon)}

\refjl{Dobrovolska I V and Tutik R S 1999} {Phys. Lett. A } {260} {10}

\refbk {Drake G W F (Ed) 1996} {Atomic, Molecular and Optical Physics Handbook} {(Woodbury: American Institute of Physics) Ch 22} 

\refjl {Elboudali F and Joulakian B 2001} {J. Phys. B: At. Mol. Opt. Phys.} {34} {4877}

\refbk {Fern\'andez F M and Castro E A 1987} {Hypervirial Theorems} {(Berlin: Springer)}

\refbk {Fern\'andez F M and Castro E A 1996} {Algebraic Methods in Quantum Chemistry and Physics} {(Boca Rat\'on: CRC)}

\refbk{Greiner W 1991} {Theoretical Physics 3: Relativistic quantum 
mechanics} {(Berlin: Springer)} 

\refjl {Ilarraza-Lomel\'i A C, Vald\'es-Mart\'inez M N,  Salas-Brito A L,  Mart\'inez-y-Ro\-mero R P, and  N\'u\~nez-Y\'epez H N 2002} {Int. J. Quantum Chem.} { 90} {195}

\refjl {Jian-You G, Xian Cheng F, and Fu-Xin X 2002} {Phys. Rev. A} {66} {062105}

\refjl {Kuang J and Lin C D 1996} {J. Phys. B: At. Mol. Opt. Phys.} {29} {L889}

\refjl {Lucha W and Sch\"oberl F F 1990} {Phys. Rev. Lett.} {23} {2733}

\refjl {Mannervik S, DeWitt D, Engstr\"om L, Lindberg J, Lindroth E, Schuch R, and Zong W 1998} {Phys. Rev. Lett.} {81} {313}

\refjl {Mart\'inez-y-Romero R P 2000} {Am. J. Phys.} {68} {1050} 
 
\refjl {Mart\'inez-y-Romero R P,  N\'u\~nez-Y\'epez H N, and  Salas-Brito H N 2000}  {J. Phys. B: At. Mol. Opt. Phys. }  {33} {L367}

\refjl {Mart\'inez-y-Romero R P, N\'u\~nez-Y\'epez H N, and  Salas-Brito A L 2001} {J. Phys. B: At. Mol. Opt. Phys.}  {34} {1261}

\refjl {Mart\'inez-y-Romero R P, N\'u\~nez-Y\'epez H N, and  Salas-Brito A L 2002} {J. Phys. B: At. Mol. Opt. Phys.}  {35} {L71}

\refjl {Mart\'inez-y-Romero R P,  N\'u\~nez-Y\'epez H N, and  Salas-Brito A L 2003} {J. Phys.\ B: At. Mol. Opt. Phys. 2003} {}  {submitted}

\refjl{Morales J 1987} {Phys. Rev. A} {36} {4101}

\refbk {Moss R E 1972} {Advanced Molecular Quantum Mechanics} {(London: Chapman and Hall)}

\refjl{M\"uller A, Phaneuf R A, Aguilar A, Gharaibeh M F, Schlachter A S, Alvarez I, Cisneros C, Hinojosa G, and McLaughlin B M  2002} {J. Phys. B: At. Mol. Opt. Phys.} {35} {L137}

\refjl {Nana Engo S G,  Kwato Njock M G,  Owono Owono L C,  Lagmago Kamta G and Motapon O 1997} {Phys. Rev. A}  {56}  {2624}

\refjl{N\'u\~nez-Y\'epez H N,  L\'opez-Bonilla J  and Salas-Brito A L 1995} {J. Phys. B: At. Mol. Opt. Phys.} {28} {L525}

\refjl {Nagy L and V\'egh L 1992} {Phys. Rev. A} {46} {284}

\refjl {Owono Owono L C,  Kwato Njock M G and  Oumaro B 2002} {Phys. Rev. A}  {66}  {052503}

\refjl {Quiney H M, Skaane H and Grant I P 1997} {J. Phys. B: At. Mol. Opt. Phys.} {30} {L829}

\refjl{Rutkowski A 1996} {Phys. Rev. A} {53} {145}

\refjl {Schippers S, M\"uller A, Ricz S, Bannister M E, Dunn G H, Bosek J, Slachter A S, Hinojosa G, Cisneros C, Aguilar A, Covington A M, Gharaibeh M F, and Phaneauf R F 2002} {Phys. Rev. Lett.}
{89} {193002}

\refjl {Schippers S, M\"uller A, Ricz S, Bannister M E, Dunn G H,  Slachter A S, Hinojosa G, Cisneros C, Aguilar A, Covington A M, Gharaibeh M F, and Phaneauf R F 2003} {Phys. Rev. A}
{67} {032702}

\refjl {Semenov S K and Cherepkov N A 2003} {J. Phys. B: At. Mol. Opt. Phys.} {36} {1409}

\refjl{Weck P, Foj\'on O A, Hanssen J, Joulakian B, and Rivarola R D 2001} {Phys. Rev. A}{63} {042709}

\refjl {West J B 2001} {J. Phys. B: At. Mol. Opt. Phys.} {34} {R45}
 
\refjl{Wong M K F and Yeh H-Y 1983a} {Phys. Rev. A}  {27} {2300}

\refjl{Wong M K F and Yeh H-Y 1983b} {Phys. Rev. A}  {27} {2305}

\refjl {Xiao-Min Ton, Lei Liu, Jia-Ming Li 1994} {Phys. Rev. A} {49} {4641}

 \vfill
 \eject
  \end